\documentclass[final,5p,times,twocolumn]{elsarticle}

\usepackage{amssymb}
\biboptions{square,sort,comma,numbers,compress}

\usepackage{lineno}

\journal{Physics Letters B}

\begin{document}
\begin{frontmatter}

\title{Effect of ground-state deformation on the Isoscalar Giant Monopole Resonance and the first observation of overtones of the Isoscalar Giant Quadrupole Resonance in rare-earth Nd isotopes}

\author[a]{M. Abdullah}
\author[a]{S. Bagchi\corref{cor}}
\cortext[cor]{Corresponding author}
\ead{sbagchi@iitism.ac.in}
\author[b]{M.N. Harakeh}
\author[c]{H. Akimune}
\author[a]{D. Das\fnref{label1}}
\fntext[label1]{Present address: TU Darmstadt and GSI Helmholtzzentrum,Germany}
\author[d]{T. Doi}
\author[e]{L. M. Donaldson}
\author[d]{Y. Fujikawa}
\author[f]{M. Fujiwara}
\author[g]{T. Furuno}
\author[h]{U. Garg}
\author[i,j]{Y.K. Gupta}
\author[h]{K.B. Howard}
\author[d]{Y. Hijikata}
\author[d]{K. Inaba}
\author[k]{S. Ishida}
\author[k]{M. Itoh}
\author[b]{N. Kalantar-Nayestanaki}
\author[a]{D. Kar}
\author[g]{T. Kawabata}
\author[c]{S. Kawashima} 
\author[a]{K. Khokhar} 
\author[c]{K. Kitamura} 
\author[f]{N. Kobayashi} 
\author[c,k]{Y. Matsuda} 
\author[k]{A. Nakagawa} 
\author[f]{S. Nakamura} 
\author[c,k]{K. Nosaka} 
\author[d]{S. Okamoto} 
\author[f]{S. Ota}
\author[a]{S. Pal}
\author[a]{R. Pramanik\fnref{label2}}
\fntext[label2]{Present address: IIT Guwahati, India}
\author[a]{S. Roy\fnref{label3}}
\fntext[label3]{Present address: IIT Roorkee, India}
\author[h]{S. Weyhmiller\fnref{label4}}
\fntext[label4]{Present address: Yale University, USA}
\author[f]{Z. Yang}
\author[l]{J.C. Zamora}

\affiliation[a]{organization={Department of Physics, IIT-ISM Dhanbad},
            postcode={Dhanbad 826004}, 
            state={Jharkhand},
            country={India}}

\affiliation[b]{organization={ ESRIG},
            addressline={University of Groningen}, 
            postcode={Groningen 9747AA}, 
            country={the Netherlands}}

\affiliation[c]{organization={Department of Physics, Konan University},
            postcode={Hyogo 658-8501}, 
            country={Japan}}  

\affiliation[d]{organization={Department of Physics, Kyoto University},
            postcode={Kyoto 606-8502},
            country={Japan}}

\affiliation[e]{organization={iThemba LABS},
            city={Somerset West},
            postcode={7129},
            country={South Africa}}

\affiliation[f]{organization={Research Center for Nuclear Physics},
            addressline={Osaka University}, 
            postcode={Osaka 567-0047},
            country={Japan}}

\affiliation[g]{organization={Department of Physics},
            addressline={Osaka University, Toyonaka}, 
            postcode={Osaka 560-0043},
            country={Japan}}

\affiliation[h]{organization={Department of Physics, University of Notre Dame},
            city={Notre Dame},
            postcode={IN 46556},
            country={USA}}

\affiliation[i]{organization={Nuclear Physics Division, Bhabha Atomic Research Centre},
            city={Mumbai},
            postcode={400085},
            country={India}}

\affiliation[j]{organization={Homi Bhabha National Institute, Anushaktinagar},
            city={Mumbai},
            postcode={400094},
            country={India}}

\affiliation[k]{organization={Cyclotron and Radioisotope Center, Tohoku University},
            postcode={Sendai 980-8578}, 
            country={Japan}}
            
\affiliation[l]{organization={Facility for Rare Isotope Beams, Michigan State University},
            city={East Lansing},
            postcode={MI 48824},
            country={USA}}

\begin{abstract}
The strength distributions of the Isoscalar Giant Monopole Resonance (ISGMR) and Isoscalar Giant Quadrupole Resonance (ISGQR) in $^{142,146-150}$Nd have been determined via inelastic $\alpha$-particle scattering with the Grand Raiden (GR) Spectrometer at the Research Center for Nuclear Physics (RCNP), Japan. In the deformed nuclei $^{146-150}$Nd, the ISGMR strength distributions exhibit a splitting into two components, while the nearly spherical nucleus $^{142}$Nd displays a single peak in the ISGMR strength distribution. A noteworthy achievement in this study is the first-time detection of overtones in the Isoscalar Giant Quadrupole Resonance (ISGQR) strength distributions within Nd isotopes at an excitation energy around 25 MeV obtained through Multipole Decomposition Analysis (MDA). 

\end{abstract}

\begin{keyword}

Isoscalar giant resonances, Compression Modes, Overtones, Ground-state deformation, Nd isotope chain.


\end{keyword}

\end{frontmatter}


\section*{Introduction}

Isoscalar giant resonances have been extensively studied in both stable and unstable nuclei~\cite{Harakeh_Book, Garg2018}. The compressional modes of the isoscalar giant resonances, namely, the Isoscalar Giant Monopole Resonance (ISGMR) and Isoscalar Giant Dipole Resonance (ISGDR) are crucial in deriving the incompressibility of finite nuclear matter ($K_{A}$)~\cite{Harakeh_Book, Stringari1982}. The incompressibility of nuclear matter ($K_{\infty}$) can, in turn, be determined from $K_{A}$ through microscopic calculations~\cite{Harakeh_Book, Garg2018}. It plays a pivotal role in the nuclear Equation-of-State (EoS), which, in turn, proves invaluable for comprehending a wide array of astrophysical phenomena, including the determination of radii and masses of neutron stars, as well as unraveling the mechanisms driving supernova explosions. These compressional modes manifest themselves as oscillations in nuclear density around a central value when the nucleus is in an excited state. 

In spherical nuclei, the strength distribution of the ISGMR does not split because of inherent directional symmetry. The influence of deformation on the Isovector Giant Dipole Resonance (IVGDR) has been extensively investigated~\cite{Berman1975}, revealing that its strength distribution splits as a result of varying oscillation frequencies along different axes in axially deformed nuclei. In deformed nuclei, the strength distribution of the isoscalar giant resonance splits into the different $K$ components, where the projection onto the symmetry axis, $K$, is a good quantum number. In the case of the ISGQR, the strength splits into three $K$ components ($K$ = 0, 1, and 2)~\cite{Kishimoto1975}. The strength of the ISGMR, characterized by the $K$ = 0 component, undergoes a splitting in deformed nuclei into two parts: one low-energy (LE) and one high-energy (HE). This splitting is attributed to the coupling of the ISGMR with the $K$ = 0 component of the ISGQR~\cite{Garg1980}. The influence of ground-state deformation on the other isoscalar modes leads to the enhancement of the widths of the strength distributions~\cite{Itoh2002, Itoh2003}. 

Isoscalar giant resonances have been studied in deformed Sm isotopes~\cite{Itoh2002,Itoh2003, Youngblood2004}, in the fission decay of $^{238}$U~\cite{Brandenburg1982}, and have been explored in a few deformed light nuclei, e.g., $^{24}$Mg~\cite{Gupta2015} and $^{28}$Si~\cite{Peach2016}. Inelastic $\alpha$-particle scattering is preferred to excite the isoscalar resonances since the $\alpha$ particle has zero spin and isospin. Our focus in this study is on measuring the strength distributions in even-A Nd isotopes, aiming to comprehend how ground-state deformation influences the distribution of the isoscalar giant resonance strength when going from spherical $^{142}$Nd to deformed $^{150}$Nd. Although isoscalar giant resonances have been studied in Nd isotopes at much lower bombarding energies~\cite{Garg1984}, the strength distributions were not extracted.

The first-order term of the transition operator of the ISGMR is a constant, while for the ISGDR, it is associated with a spurious center-of-mass motion. Consequently, only the second-order terms lead to intrinsic collective excitations of the nucleus for these modes of giant resonances~\cite{Harakeh_Book}, whereas the ISGQR is described by the first term of the transition operator. Except for specific cases, such as the ISGMR and ISGDR, giant resonances corresponding to higher-order terms, often referred to as overtones, have remained elusive. Notably, in Ref.~\cite{Hunyadi2003}, an overtone mode in the quadrupole strength distribution has been found in the proton-decay of $^{208}$Pb for the first time. The excitation energy and the width of this ``$L=2$" mode were 26.9 $\pm$ 0.7~MeV and $\Gamma$ = 6.0 $\pm$ 1.3 MeV, respectively. An overtone mode of ``$L=2$" character was also hinted at in the neutron decay of $^{208}$Pb~\cite{Hunyadi2007}. Continuum Random-Phase-Approximation (RPA) calculations based on a partially self-consistent semi-microscopic approach have predicted the existence of overtone modes of quadrupole nature with the calculated centroid energy higher than 30 MeV~\cite{Gorelik2004}. The overtone modes have a compressional character that is used to determine the nuclear incompressibility by measuring the centroid energies of the strength distributions. Hence, the higher-order mode in the ISGQR (4$\hbar \omega$ excitation) refers to another compression mode in addition to the ISGMR and ISGDR.

In this letter, we report the experimental results demonstrating the effect of nuclear ground-state deformation on the ISGMR in $^{142,146-150}$Nd,  where the deformation increases from the spherical $^{142}$Nd ($\beta_{2}$ = 0.0916(8)) to deformed $^{150}$Nd ($\beta_{2}$ = 0.285(3))~\cite{NNDC}. Distinctive signatures of overtones in $^{142,146-150}$Nd, having excitation energies between 25 $-$ 30 MeV with $L$ = 2 character, have been unveiled by employing Multipole Decomposition Analysis (MDA). 

\section*{Experimental Setup}

The experiment was performed at RCNP, Osaka University, Japan. A halo-free beam of $\alpha$ particles, having a total energy of 386~MeV, accelerated through the AVF and ring cyclotrons, impinged on self-supporting and enriched $^{142,146,148,150}$Nd targets. These targets had areal densities of approximately 5 mg/cm$^2$. The beam current ranged from 0.1 to 10 nA. The energy resolution was around 175 keV, which was sufficient for investigating the giant resonances. Inelastically-scattered $\alpha$ particles were momentum analyzed in the high-resolution Grand Raiden (GR) spectrometer~\cite{Fujiwara1999}. Subsequently, they were transported to the final focal plane for detection. The focal-plane detection system was composed of a pair of Multi-Wire Drift Chambers (MWDCs) for vertical- and horizontal-position measurements, along with two plastic-scintillator detectors for particle identification and triggering the data acquisition. The vertical-focusing mode of the GR spectrometer facilitates focusing the true events originating from scattering off the target in a narrow band along the vertical plane whereas events stemming from instrumental background are over- or under-focused in the vertical plane. Subtraction of the events in the off-median focal-plane positions from the events in the median focal-plane position facilitates an efficient removal of the instrumental background~\cite{Itoh2003, Fujiwara_Book}. The inelastically scattered $\alpha$ particles were measured at forward angles, 0$^{\circ}$ $\leq$ $\rm\theta_{Lab}$ $\leq$ 10$^{\circ}$, where the angular distributions are characteristic of different multipolarities. To calibrate the energy spectra, runs with a $^{24}$Mg target (thickness 2.5 mg/cm$^2$) were performed. The calibration was made using the low-excitation-energy peaks of $^{24}$Mg obtained from inelastic $\alpha$ scattering.

\section*{Data Analysis}

\begin{figure}
\includegraphics[width=1.0\linewidth]{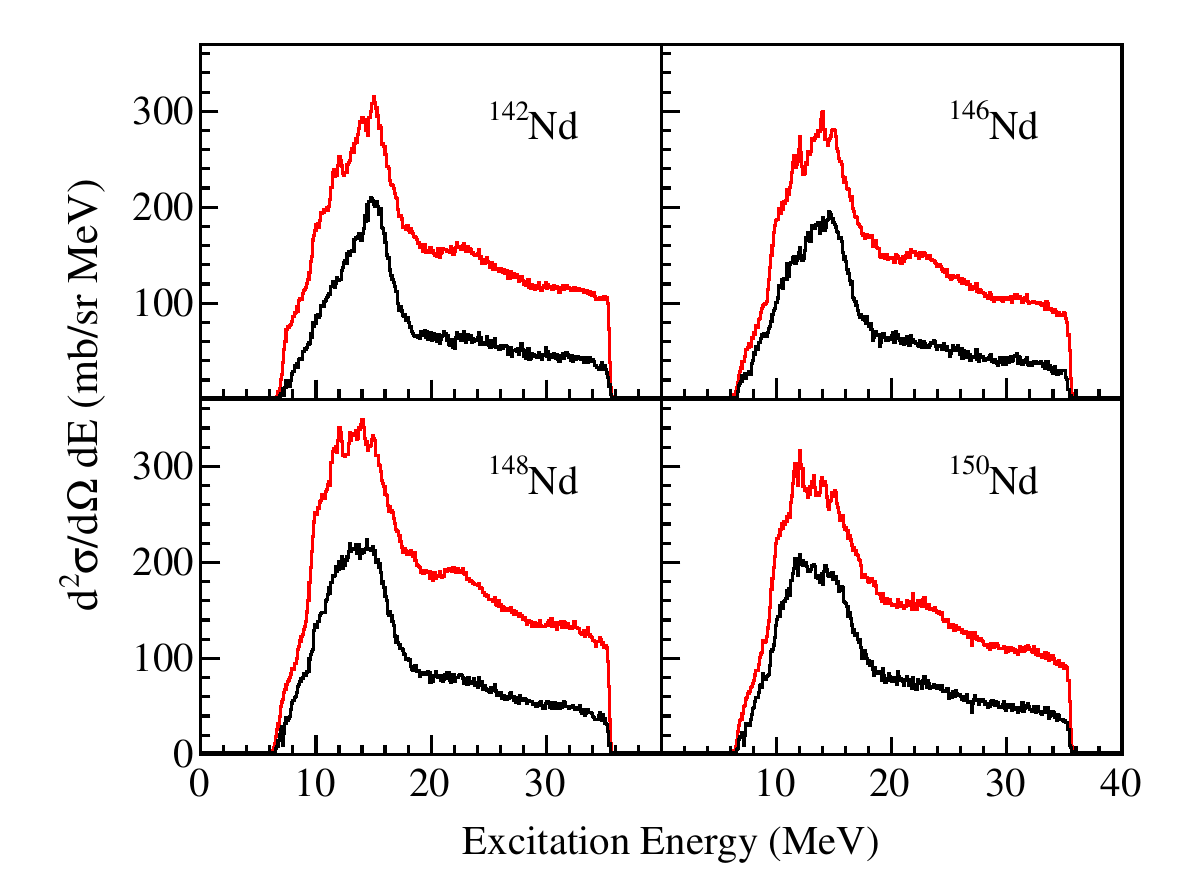}
\caption{Measured double-differential cross-section spectra from $^{142,146,148,150}$Nd($\alpha$,$\alpha'$)
at $\theta_{\rm Lab}$ = 0.75$^{\circ}$ after particle identification, subtraction of the instrumental background, ion-optical correction, and subtraction of contaminants are shown in black histograms whereas the red histograms show the spectra before instrumental-background and contaminants subtraction. The peaks in the red histograms are from oxygen contamination.} 
\label{Zerodegexcitationenergy}
\end{figure}

The excitation-energy spectra of the Nd isotopes are extracted after particle identification, instrumental background subtraction, ion-optical corrections, and subsequent calibration in the offline analysis. Each Nd isotope has substantial amounts of hydrogen and oxygen contamination. In the case of hydrogen contamination, only the elastic channel must be accounted for, as the elastic scattering off the protons has an enormous cross-section compared to the inelastic excitation of the target nucleus. Therefore, by removing the affected data points the hydrogen contamination can be eliminated. On the other hand, to eliminate the oxygen contamination, we used the high-resolution $^{16}$O($\alpha,\alpha')$ measurements performed at the same beam energy at RCNP~\cite{ItohPriv}. After aligning the kinematics with the measured excitation energies of the Nd inelastic spectra, the oxygen excitation-energy spectra are smeared with the experimental resolution. These spectra are then scaled by the ratio of the integrals of the prominent oxygen peaks in the Nd excitation-energy spectra to those in the kinematically transformed oxygen spectra. Finally, the scaled oxygen spectra are subtracted from the Nd excitation-energy spectra to remove the oxygen contamination. In Fig.~\ref{Zerodegexcitationenergy}, the raw Nd excitation-energy spectra are shown in red, whereas the black histograms represent Nd excitation-energy spectra after the subtraction of the contaminants and instrumental background at $\theta_{\rm Lab}$ = 0.75$^{\circ}$. The contribution of the instrumental background is the largest near $\theta_{\rm GR}$ = 0$^{\circ}$ due to the elastic scattering of $\alpha$ particles off the beamline elements. A detailed description of the calibration and the data-reduction processes is provided in Ref.~\cite{Abdullah2025}.

The GR spectrometer was set to different angles from 0$^{\circ}$ to 9.5$^{\circ}$ to measure the inelastic scattering data for $^{142,146,148,150}$Nd over the angular range 0$^{\circ}$ $\leq$ $\rm{\theta_{Lab}}$ $\leq$ 10${^\circ}$. At the final dispersive focal plane, the acceptance of the spectrometer ranged from 10 to 30 MeV. For each angular setting of the GR spectrometer, a precise energy calibration was performed using the inelastic $\alpha$ scattering off $^{24}$Mg and comparing the excitation-energy spectrum with well-known low-lying states in $^{24}$Mg obtained from high-resolution $^{24}$Mg($\alpha$,$\alpha'$) data~\cite{KawabataPriv}. The 0$^{\circ}$ spectra, where the ISGMR cross sections are maximal, are illustrated in Fig.~\ref{Zerodegexcitationenergy}, where the resonance visibly broadens with increasing deformation from $^{142}$Nd to $^{150}$Nd. 

To extract the strength distributions of the giant resonances through MDA, it is imperative to obtain the Optical-Model Parameters (OMPs). To obtain the OMPs, elastic scattering on $^{142}$Nd was performed over a broader angular range of 3.5$^{\circ}$ to 20.5${^\circ}$. These OMPs are employed in the framework of the Distorted-Wave Born Approximation (DWBA) calculation using the coupled-channel code CHUCK3 \cite{CHUCK3}. We used the Woods-Saxon type of optical potential for both real and imaginary parts. The OMPs are derived through a $\chi^2$ fitting of the $\alpha$ elastic-scattering data on $^{142}$Nd over an angular range of 3.5$^{\circ}$ to 20.5${^\circ}$. Since the $\alpha$ elastic-scattering data on $^{146,148,150}$Nd were unavailable due to limited beam time, we employed the OMPs obtained from $^{142}$Nd data in the DWBA calculations involving $^{146,148,150}$Nd. The use of OMPs from a nearby nucleus or isotope has a negligible effect on the extraction of the strength distributions of giant resonances~\cite{Li2010, Howard2020, Bagchi2015, Howard2020_1}. Furthermore, OMPs derived for a spherical nucleus have been used to determine the strength distributions of deformed nuclei without altering their expected characteristics~\cite{Itoh2002, Itoh2003}. Utilizing the obtained OMPs and the known $B(E2)$ and $B(E3)$ values, the angular distributions for the states at 1.575 MeV ($J^{\pi}$ = 2$^{+}$) and 2.083 MeV ($J^{\pi}$ = 3$^{-}$) were calculated within the DWBA framework and they are in agreement with the experimental data~\cite{Abdullah2025, Abdullah2024, Das2021}.

While the isoscalar giant resonances are excited by the scalar, isoscalar $\alpha$ probe, it should be noted that the Coulomb excitation can lead to contributions from the isovector modes. Consequently, an $\alpha$ particle can excite the IVGDR mode with a non-negligible cross section. The contribution of the IVGDR has to be subtracted prior to performing MDA to extract the strength distribution of the isoscalar giant resonances~\cite{Patel2014, Li2010, Howard2020}. The form factors for the IVGDR are calculated using the Goldhaber-Teller model~\cite{Satchler1987, Krasznahorkay1991, Krasznahorkay1994} and the exhausted sum rules are estimated using the available photo-neutron data~\cite{Plujko2018} for Nd isotopes. These estimates facilitate the determination of the contribution from the IVGDR at each excitation-energy interval. The angular distributions for ($\alpha$, $\alpha'$) scattering over the excitation-energy range 10 $-$ 30 MeV are extracted in 1-MeV-wide bins to mitigate the statistical fluctuations. Subsequently, these angular distributions are then decomposed into a linear combination of the calculated DWBA angular distributions, as demonstrated in Eq.~\ref{MDA_eqn}.
\begin{equation}
\frac{d^2\sigma^{\rm exp}(\theta_{\rm CM}, E_{\rm x})}{d\Omega dE} = \sum_{L} A_{L}(E_{\rm x}) \frac{d^2\sigma^{\rm DWBA}(\theta_{\rm CM}, E_{\rm x})}{d\Omega dE}
\label{MDA_eqn}
\end{equation}

\begin{figure*}
\centering
\includegraphics[width=15cm,height=7.5cm,keepaspectratio]{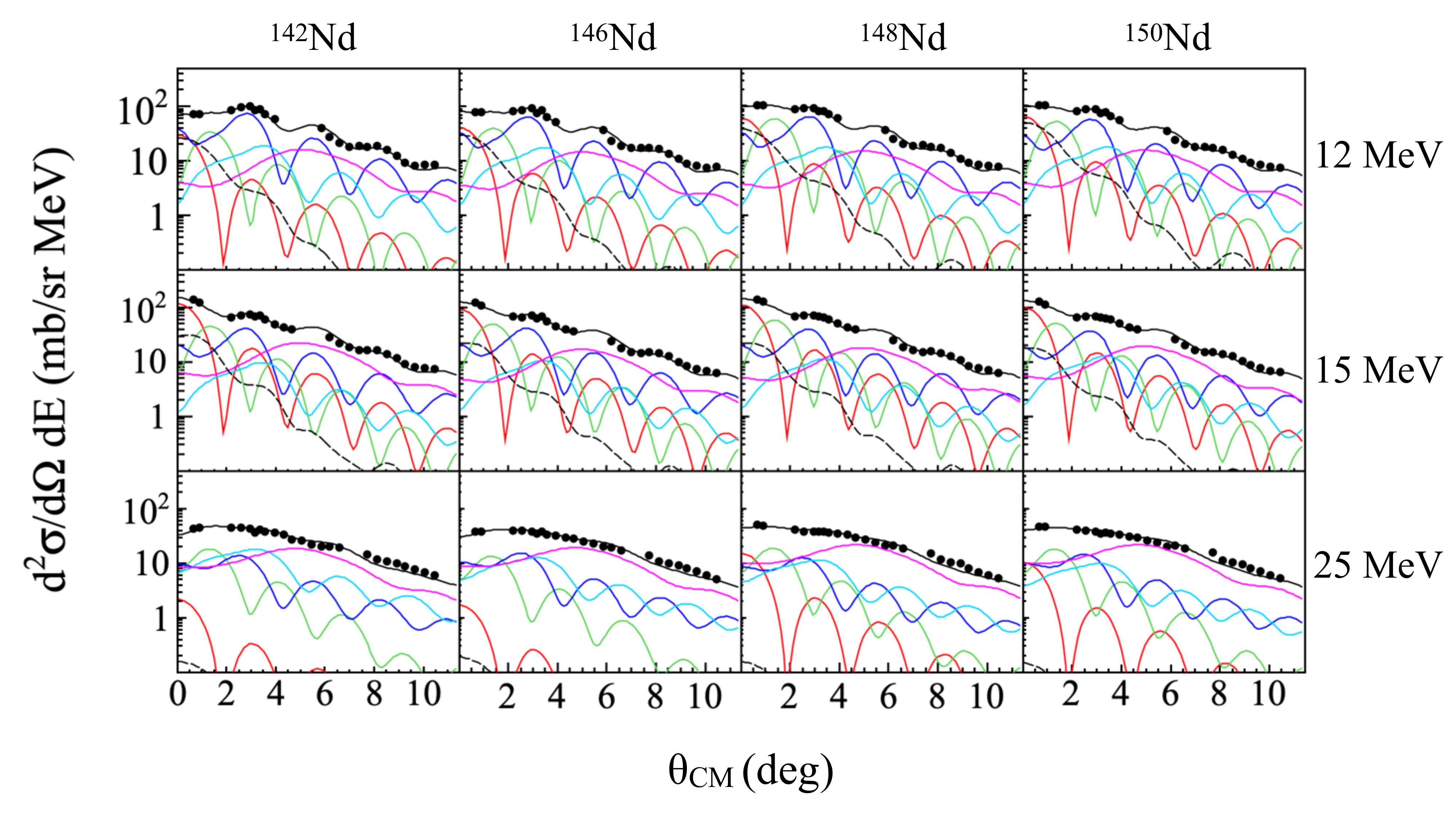}
\caption{Multipole-decomposition analyses for $^{142,146,148,150}$Nd are shown for excitation-energy bins centered at 12 MeV, 15 MeV, and 25 MeV. Total fits to the data points are represented by solid black lines. Fitted contributions are also shown from the isoscalar monopole (red), dipole (green), quadrupole (blue), octupole (cyan), and higher multipole modes (magenta). The contributions from the IVGDR, determined through known photo-neutron cross-section data and the Goldhaber-Teller model, are also depicted by dashed lines.} 
\label{MDAfit}
\end{figure*}

The $A_L(E_{\rm{x}})$ coefficients represent the fraction of the Energy-Weighted Sum Rule (EWSR) for the multipolarity $L$ for a particular energy bin~\cite{Harakeh_Book}. Our analysis encompasses multipoles up to $L$ = 7 in the fitting process. The DWBA calculations are carried out using the CHUCK3 code considering 100\% exhaustion of the EWSR for each multipole at each energy bin. The coefficients $A_L(E_{\rm x})$ for different multipoles are determined by $\chi^{2}$ minimization. The cross sections decrease and their angular distributions become featureless at larger angles, while the physical continuum dominates at higher excitation energies. Consequently, the strengths could only be reliably extracted up to $L$ = 3. For values beyond $L$ = 3, we summed the angular distributions for $L\ge4$ before fitting the experimental data. Once the fitting parameters are obtained, the error bar in each parameter is then determined after fixing the other parameters to raise the confidence level by 68\%. Typical results of the MDA are presented in Fig.~\ref{MDAfit}. Transition potentials are obtained from collective transition densities following the methods described in Refs.~\cite{Harakeh_Book, Harakeh1981}. The transition densities are obtained from the sum-rule approach and expressed in terms of ground-state densities. Such a model is used to obtain the form factors for DWBA analysis. Using the coefficients $A_L$, the strength distributions for the ISGMR and ISGQR are obtained from Eq.~\ref{Eqn_Strength_GMR} and Eq.~\ref{Eqn_Strength_GQR}, respectively~\cite{Harakeh_Book, Harakeh1981}.

\begin{equation}
 {S_{0}(E_{\rm x})} = \frac{\hbar^2}{2m}\frac{ Z^2}{A}\frac{<r^2>}{E_{\rm x}}A_{0}(E_{\rm x})
    \label{Eqn_Strength_GMR}
\end{equation}

\begin{equation}
   {S_{2}(E_{\rm x})} = \frac{50\hbar^2}{8 \pi m}\frac{ Z^2}{A}\frac{<r^2>}{E_{\rm x}}A_{2}(E_{\rm x})
    \label{Eqn_Strength_GQR}
\end{equation}

where $m$ is the nucleon mass, $A$ is the mass number, and $<r^2>$ is the rms value of the ground-state density. The extracted ISGMR and ISGQR strength distributions of the Nd isotopes are shown in Fig.~\ref{E0_strength} and Fig.~\ref{E2_strength}, respectively. The strength distributions of other multipoles are reported in Refs.~\cite{Abdullah2024, Abdullah2025}. A preliminary report can be found in Ref.~\cite{Abdullah2022}.

\section*{Results and discussions}

\begin{figure}[htb]
\includegraphics[width=1.0\linewidth]{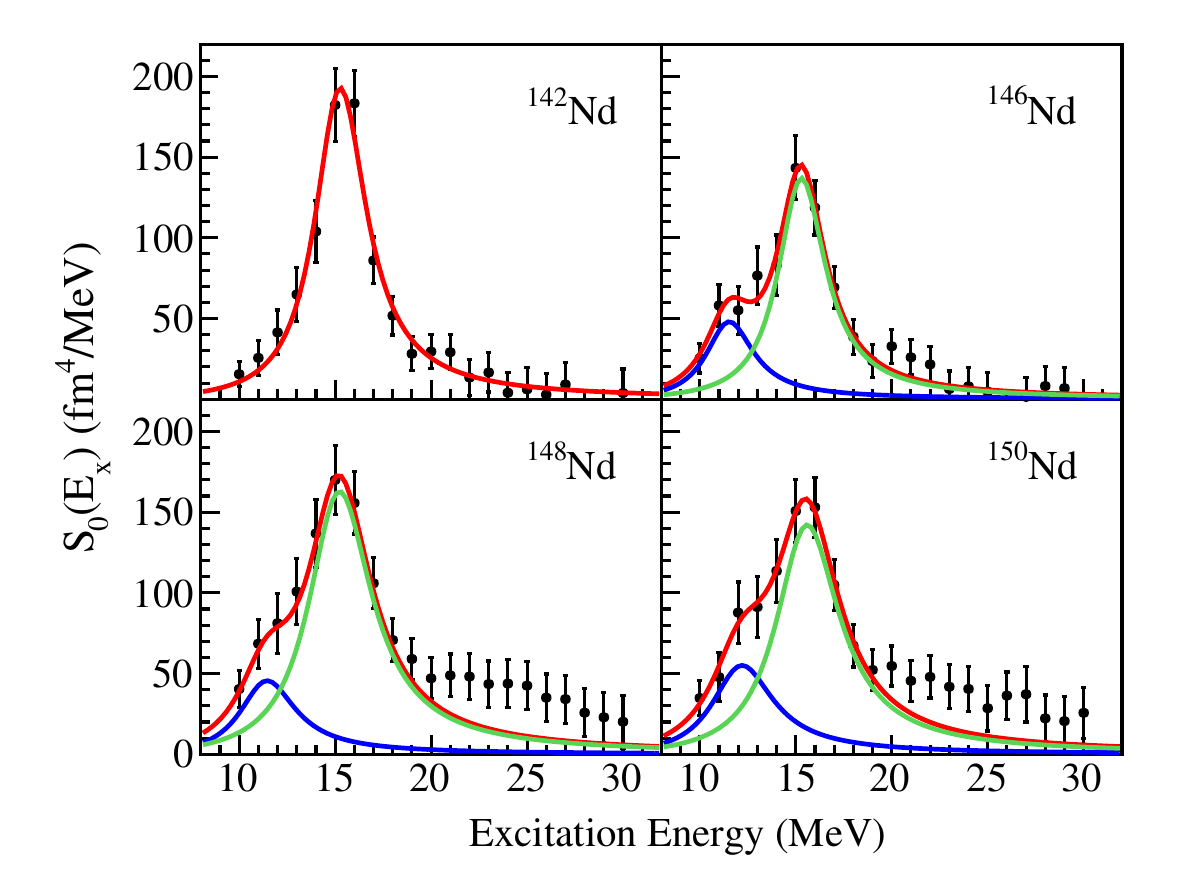}
\caption{The monopole strength distributions for $^{142-150}$Nd were obtained from the MDA. The red lines show the results of single-Lorentzian fitting in $^{142}$Nd (upper left) and of double-Lorentzian fitting in $^{146}$Nd (upper right), $^{148}$Nd (lower left), and $^{150}$Nd (lower right). In deformed Nd isotopes, the low- and high-energy components of the ISGMR are indicated by blue and green lines, respectively.}
\label{E0_strength}
\end{figure}

The ISGMR strength is expected to split into two components due to the coupling to the $K$ = 0 component of the ISGQR during the transition from spherical $^{142}$Nd to prolate-shaped $^{146,148,150}$Nd isotopes. The extracted ISGMR strength distribution for $^{142}$Nd has been fitted with a single Lorentzian function (see the form of the Lorentzian function in Ref.~\cite{Berman1975}).  For $^{146,148,150}$Nd, a double-Lorentzian function has been used since the ISGMR strength distribution splits due to the deformation. We have used the fitting range of 10 $-$ 18 MeV for $^{142,148}$Nd and 10 $-$ 19 MeV for $^{146,150}$Nd. The results from the fitting are shown in Fig.~\ref{E0_strength} and fitting parameters are shown in Tab.~\ref{Tab1}. The error bars in the strength distributions are estimated from the errors obtained through the $\chi^{2}$ minimization performed in the MDA. These errors are not statistical and hence the choice of the fitting range affects the errors of the Lorentzian fitting parameters. Since $^{142}$Nd is spherical, only one component exists, whereas, in the case of deformed $^{146,148,150}$Nd nuclei, the peak position of the LE component matches with the $K$ = 0 component of the ISGQR within the error bars (see Tab.~\ref{Tab1} and Tab.~\ref{Tab2} and also refer to Fig. 1 of Ref.~\cite{Kvasil2016}), showing the onset of monopole-quadrupole coupling as the deformation increases. Surprisingly, our findings reveal a comparatively lower relative strength for the monopole mode at high excitation energies (above 20 MeV) in contrast to the results reported in Ref.~\cite{Itoh2002} for Sm isotopes. The EWSR fractions are obtained by integrating $E_{\rm x}S_{0}(E_{\rm x})$ over the energy 10 $-$ 22 MeV, where $S_{0}(E_{\rm x})$ is estimated from the Lorentzian fits. The EWSR of the LE peak increases with the increase in deformation, whereas the EWSR of the HE peak remains the same with a small decrease in $^{146}$Nd. A near-constant monopole strength is observed at high excitation energies in $^{148}$Nd and $^{150}$Nd, likely originating from the physical continuum, such as knock-out reactions and quasi-free processes~\cite{Garg2018}. Furthermore, it should be remarked that although subtraction of the oxygen contribution may affect the strength distribution, it does not alter the peak position or width.

\begin{table*}[htb]
\centering
\begin{tabular}{l|ccc|ccc}
\hline
Isotope & E$_{\rm LE}$(MeV) &$\Gamma_{\rm LE}$(MeV) & \%EWSR$_{\rm LE}$ & E$_{\rm HE}$(MeV) & $\Gamma_{\rm HE}$(MeV) & \%EWSR$_{\rm HE}$\\ %
 \hline
 $^{142}$Nd \rule{0pt}{4ex}  & $-$ & $-$ & $-$ & 15.3 $\pm$ 0.1 & 3.3 $\pm$ 0.2 & 103.9$^{+10.9}_{-14.3}$\\
 
$^{146}$Nd \rule{0pt}{4ex}  & 11.5 $\pm$ 0.4 & 3.0 $\pm$ 0.8 & 17.6$^{+9.2}_{-7.4}$ & 15.3 $\pm$ 0.2 & 3.0 $\pm$ 0.3
 & 67.8$^{+12.0}_{-11.2}$\\

$^{148}$Nd \rule{0pt}{4ex}  & 11.5 $\pm$ 0.5 & 3.5 $\pm$ 1.2 & 19.1$^{+12.7}_{-9.7}$ & 15.2 $\pm$ 0.2 & 4.3 $\pm$ 0.4 & 108.0$^{+15.9}_{-15.2}$\\

$^{150}$Nd \rule{0pt}{4ex}  & 12.2 $\pm$ 0.4 & 3.8 $\pm$ 1.0 & 27.6$^{+13.3}_{-10.9}$ & 15.6 $\pm$ 0.2 & 4.0 $\pm$ 0.4 & 90.9$^{+14.4}_{-13.7}$ \vspace{0.1em}\\

\hline
\end{tabular}
\caption{The parameters obtained from the Lorentzian fitting of the strength distributions of the ISGMR in Nd isotopes are presented. The centroid and the width of the low-energy (LE) component are depicted as E$_{\rm LE}$ and $\Gamma_{\rm LE}$, respectively, while those of the high-energy (HE) component are depicted as E$_{\rm HE}$ and $\Gamma_{\rm HE}$, respectively.  The \%EWSRs are calculated from the fits to the strength distributions in Fig.~\ref{E0_strength}. The errors in \%EWSRs include the uncertainties stemming from the parameters obtained through the Lorentzian fitting process. They are estimated over the range of E$_{\rm x}$ = 10 $-$ 22 MeV for both the LE (single Lorentzian represented by blue lines in Fig.~\ref{E0_strength}) and HE (single Lorentzian represented by green lines in Fig.~\ref{E0_strength}) components for $^{146,148,150}$Nd, whereas only the HE component is present for $^{142}$Nd. The error bars listed for the peak and the width are consistent with a 68\% confidence interval. }
\label{Tab1}
\end{table*}

\begin{figure}[htb]
\includegraphics[width=0.9\linewidth]{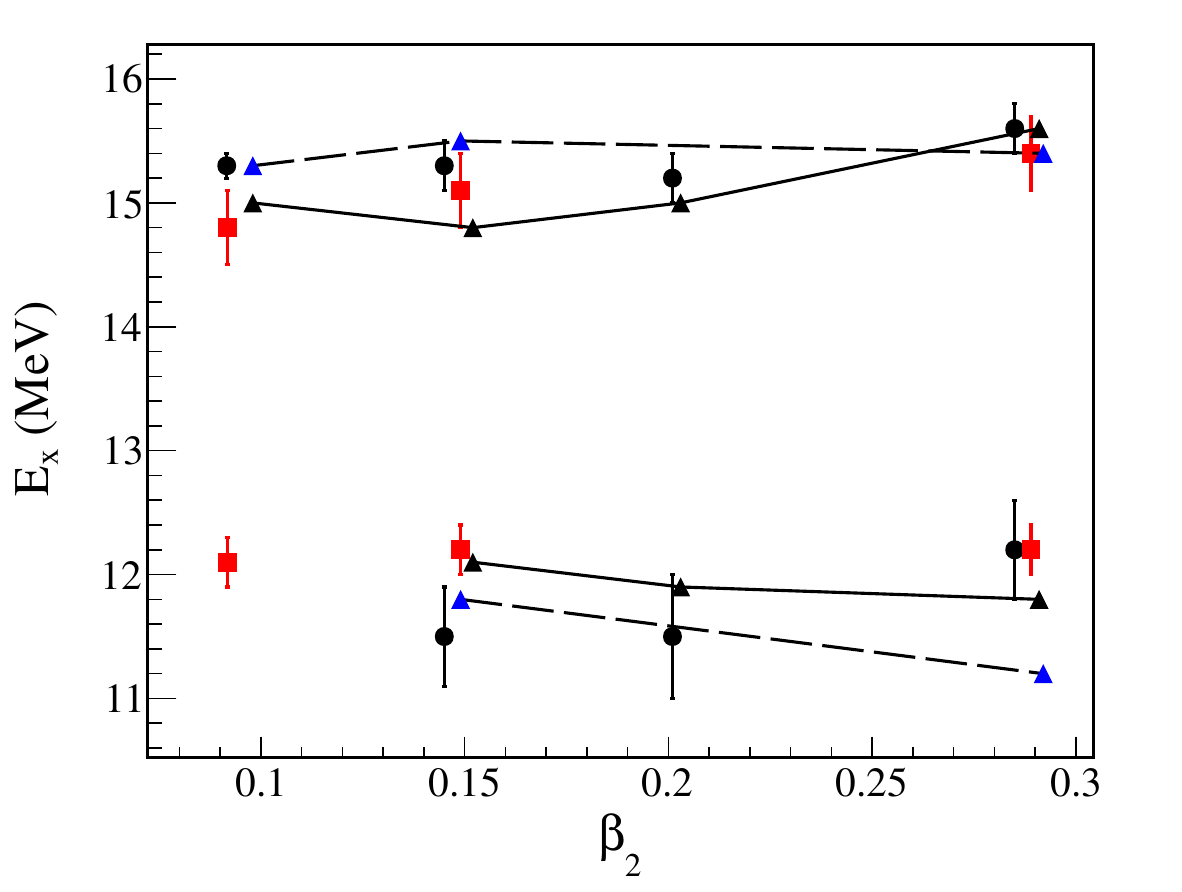}
\caption{The peak positions of the LE and HE components of the ISGMR strength distributions are shown for Nd isotopes as functions of the deformation parameter and compared with other data. The black-filled circles represent the present experimental findings. The results from Ref.~\cite{Garg1984} are denoted by red-filled squares. QRPA calculations with the SkM* Skyrme energy density functional~\cite{Yoshida2013} are shown with black-filled triangles. Self-consistent QRPA calculations with SV-bas and SkP Skyrme energy density functionals~\cite{Kvasil2016} are shown with blue-filled triangles. The solid and dashed lines are guides connecting the theoretical results. The deformation parameters (central values) are obtained from Ref.~\cite{NNDC}. The abscissae of the data points, excluding the present experimental data, are arbitrarily shifted for better visualization.} 
\label{GMRpeak}
\end{figure}

Garg et al.~\cite{Garg1984} performed $\alpha$-particle inelastic scattering on $^{142,146,150}$Nd at a beam energy of 129~MeV. The excitation-energy spectra, after subtraction of a smooth continuum background, were fitted with a double-Gaussian function for the LE and HE components even for $^{142}$Nd. However, in this work, the strength distributions of the giant resonances were not extracted. In Fig.~\ref{GMRpeak}, the peak positions extracted from the Lorentzian fits from the present work are compared with those from the Gaussian fits obtained from Ref.~\cite{Garg1984} and they are in agreement within the error bars. Giant resonances in Nd isotopes have been studied by employing the Quasiparticle-Random-Phase-Approximation (QRPA) with Skyrme energy density functional, namely SkM*~\cite{Yoshida2013}. The peak positions, extracted from the Lorentzian fitting of the calculated strength distributions, are also shown in Fig.~\ref{GMRpeak}, demonstrating the agreement with our findings within the error bars. Additionally, Kvasil et al.~\cite{Kvasil2016} also performed self-consistent QRPA calculations to assess the strength distribution in $^{142}$Nd, $^{146}$Nd, and $^{150}$Nd. These QRPA calculations utilized two different Skyrme forces, one with a large (SV-bas) and one with a small (SkP) nuclear incompressibility, resulting in consistent centroid energies for the ISGMR in Nd isotopes. 

\begin{figure}[htb]
\includegraphics[width=1.0\linewidth]{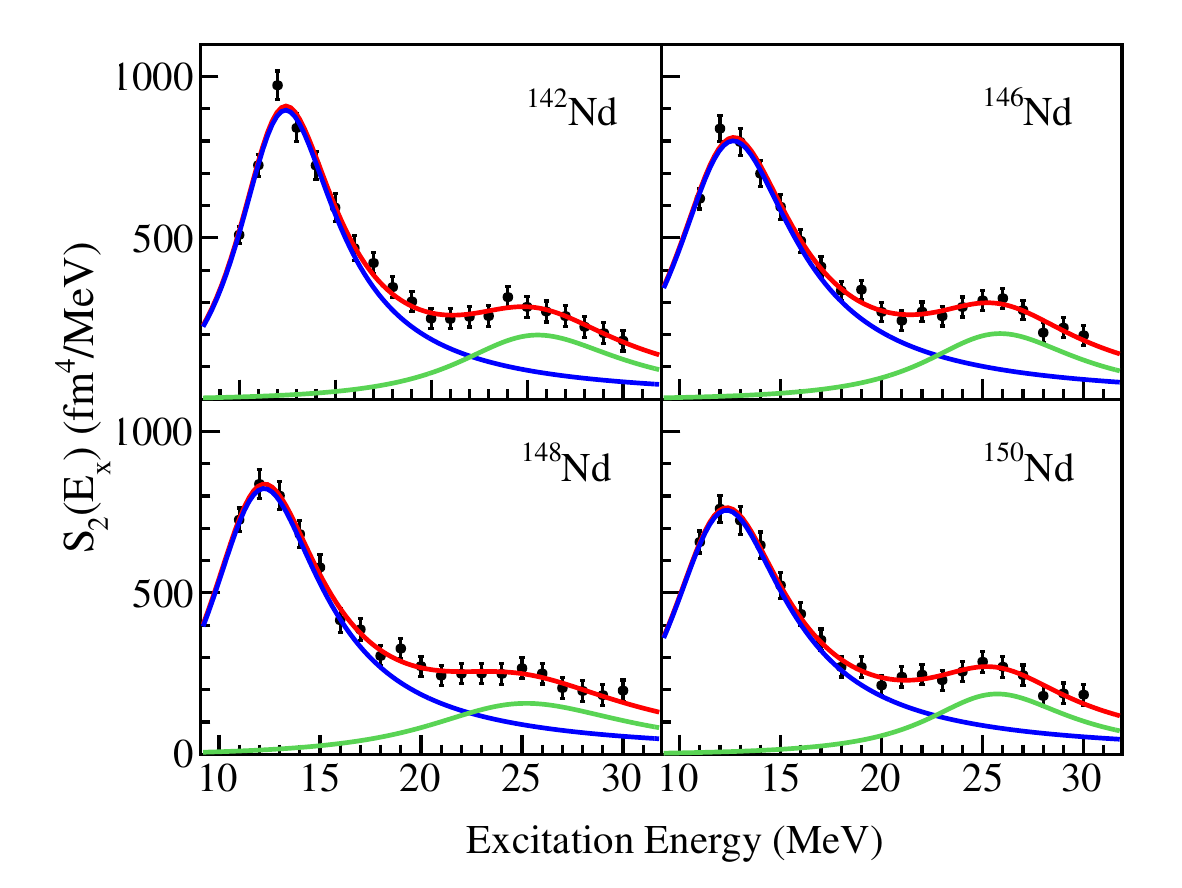}
\caption{The quadrupole strength distributions for $^{142-150}$Nd isotopes were obtained from the MDA. The overall fits, employing a double-Lorentzian function, are represented by red lines. The maintone modes of the quadrupole resonances are depicted using blue lines, while the overtone modes are illustrated with green lines.}
\label{E2_strength}
\end{figure}

The ISGQR strength distributions for the Nd isotopes, derived through MDA, are presented in Fig.~\ref{E2_strength}. The strength distributions are fitted with a double-Lorentzian function over the energy range 10 $-$ 30 MeV. The fitting results are listed in Tab.~\ref{Tab2}. The ISGQRs exhibit a maintone associated with the operator having the form $\sum_{i} r_{i}^{2}Y_{2}$ and the frequency of this mode is 2$\hbar \omega$. With increasing deformation, the ISGQR strength distribution splits into three components, namely, $K$ = 0, 1, and 2. However, this phenomenon is reflected in a broadened width of the strength distribution~\cite{Yoshida2013}. This is also evident from the widths extracted from the Lorentzian fitting (see Tab.~\ref{Tab2}). 

\begin{table*}[htb]
\centering
\begin{tabular}{c|ccc|ccc}
\hline
 &\multicolumn{1}{c}{Maintone Mode}&\multicolumn{5}{c}{Overtone Mode}\\
 \hline

Isotope & LE$_{\rm GQR}$(MeV) & $\Gamma_{\rm LE}$(MeV) & \%EWSR$_{\rm LE}$ & HE$_{\rm GQR}$(MeV) & $\Gamma_{\rm HE}$(MeV) & \%EWSR$_{\rm HE}$\\ 
 \hline
 $^{142}$Nd \rule{0pt}{4ex}  & 12.4 $\pm $ 0.1 & 6.3 $\pm$ 0.2 & 109.6$^{+3.5}_{-3.5}$ & 25.5 $\pm$ 0.4 & 10.5 $\pm$ 1.3 & 78.8$^{+7.2}_{-8.0}$\\
 
 $^{146}$Nd \rule{0pt}{4ex}  & 12.7 $\pm$ 0.1 & 7.1 $\pm$ 0.2 & 103.9$ ^{+3.2}_{-3.3}$ & 25.9 $\pm$ 0.4 & 9.3 $\pm$ 1.1 & 77.8$^{+7.1}_{-7.7}$\\

 $^{148}$Nd \rule{0pt}{4ex}  & 12.2 $\pm$ 0.1 & 6.7 $\pm$ 0.2 & 104.2$^{+4.0}_{-4.1}$ & 25.2 $\pm$ 0.6 & 12.3 $\pm$ 2.0 & 67.1$^{+7.2}_{-8.4}$\\

 $^{150}$Nd \rule{0pt}{4ex}  & 12.3 $\pm$ 0.1 & 6.9 $\pm$ 0.2 & 97.5$^{+2.5}_{-2.6}$ & 25.7 $\pm$ 0.4 & 8.7 $\pm$ 1.1 & 70.7$^{+7.7}_{-8.4}$ \vspace{0.1em}\\
 
\hline
\end{tabular}
\caption{The parameters from the Lorentzian fitting of the ISGQR strength distributions in Nd isotopes are listed. The peak positions and the widths of the maintone modes, illustrated by single Lorentzian functions represented by blue lines in Fig.~\ref{E2_strength}, are depicted as $\rm LE_{\rm GQR}$ and $\Gamma_{\rm LE}$, respectively, while that of the overtone modes, illustrated by single Lorentzian functions represented by green lines in Fig.~\ref{E2_strength},  are depicted as $\rm HE_{\rm GQR}$ and $\Gamma_{\rm HE}$, respectively. The \% EWSRs are computed from the fits to the strength distributions in Fig.~\ref{E2_strength} within the energy intervals of 10 $-$ 16 MeV for the maintone component and 20 $-$ 30 MeV for the overtone peak. A 68\% confidence interval has been achieved in the errors related to the peak positions and the widths. The errors in the \%EWSR encompass the uncertainties derived from the parameters obtained through the Lorentzian-fitting process.}
\label{Tab2}
\end{table*}

The first evidence for a high-lying resonance around 27 MeV exhibiting $E2$ character was reported in the proton decay of the ISGDR in $^{208}$Pb~\cite{Hunyadi2003}. It leads to the third compression mode in addition to the ones related to the ISGMR and ISGDR. The frequency of this overtone is 4$\hbar\omega$. In the ISGQR strength distribution in Fig.~\ref{E2_strength}, which emerged from the MDA analysis, an $L$ = 2 bump is visible at high excitation energy ($\sim$ 25 MeV), suggesting the existence of the overtone mode of the ISGQR. We have used the operator having the form $\sum_{i} r_{i}^{2}Y_{2}$ for obtaining the strength distributions for both maintone and overtone quadrupole modes. The parameters from the Lorentzian fits are detailed in Tab.~\ref{Tab2}. The EWSR fractions are obtained by integrating $E_{\rm x}S_{2}(E_{\rm x})$ over the energy range 10 $-$ 16 MeV for the maintone mode and over the energy range 20 $-$ 30 MeV for the overtone mode. The strength $S_{2}(E_{\rm x})$ is estimated from the Lorentzian fits. It is noteworthy that the overtone mode of quadrupole character was not observed in inelastic deuteron and $\alpha$ scattering on $^{208}$Pb~\cite{Uchida2003, Patel2014} and $\alpha$ scattering on Sm isotopes~\cite{Itoh2002, Itoh2003} performed using the GR spectrometer at RCNP. However, some indications of quadrupole overtone modes were seen as broad peaks above 20 MeV in Ref.~\cite{Gupta2018} for $^{90,92}$Zr and $^{92}$Mo. Furthermore, as mentioned above, an overtone mode in the quadrupole strength distribution has been found in the proton-decay of $^{208}$Pb~\cite{Hunyadi2003}. Therefore, further experimental evidence of overtones in other nuclei is essential to corroborate these findings.

\section*{Summary}

In summary, we have measured the inelastic $\alpha$ scattering on even-A Nd isotopes ($^{142,146,148,150}$Nd) using the Grand Raiden spectrometer at RCNP at a beam energy of 386 MeV and at forward angles. After particle identification, instrumental background subtraction, ion-optical correction, and subtraction of hydrogen and oxygen contaminants, we performed MDA analysis using DWBA angular distributions where the form factors were obtained from the ground-state densities. The ISGMR strength distributions clearly show a splitting as the ground-state shape deformation increases. The obtained peak positions at both low and high energies are in agreement with the previous measurement at Texas A\&M University~\cite{Garg1984}. The results also agree well with QRPA calculations~\cite{Yoshida2013, Kvasil2016}. The analysis also encompasses the extraction of the  ISGQR strength distributions through MDA analysis. A distinct signature of an overtone mode in the ISGQR has emerged, implying the existence of a third category of compression modes in addition to the ISGMR and ISGDR. Further exploration in this context holds the potential to firmly establish the $L$=2 compression mode. For a comprehensive understanding of our analytical methodology, which includes the strength distributions for other multipoles, readers are referred to the detailed analysis procedures outlined in Refs.~\cite{Abdullah2024, Abdullah2025}.

\section*{Acknowledgments}

The authors gratefully thank the dedicated staff of the RCNP Ring Cyclotron Facility for providing a high-quality, halo-free $\alpha$ beam. The local support provided by RCNP during the experiment is deeply acknowledged. SB is grateful to A. Tamii for the help in the procurement of the targets. SB is indebted to the Science and Engineering Research Board (SERB), India (Grant No. SRG/2021/000827) and the Faculty Research Scheme at IIT (ISM) Dhanbad (Grant No. FRS(154)/2021-2022/Physics) for the financial support. SB is also thankful to GSI Helmholtzzentrum, Germany for the travel support for participating in the experiment. MA is thankful to MHRD, Government of India, for the financial support. This work was supported in part by the U. S. National Science Foundation (Grants No. PHY-1713857 and No. PHY-2310059). The isotopes used in this research were supplied by the United States Department of Energy Office of Science by the Isotope Program in the Office of Nuclear Physics. 


\end{document}